# Physical properties, electronic structure, and strain-tuned monolayer of the weak topological insulator RbTi$_3$Bi$_5$ with Kagome lattice

Ying Zhou[1, 2, #], Long Chen[1, 2, #], Xuecong Ji[1, 2, #], Chen Liu[3], Ke Liao[1, 2], Zhongnan Guo[4], Jia'ou Wang[3], Hongming Weng[1, 2, 5, *], Gang Wang[1, 2, 5, *]

[1] Beijing National Laboratory for Condensed Matter Physics, Institute of Physics, Chinese Academy of Sciences, Beijing 100190, China

[2] University of Chinese Academy of Sciences, Beijing 100049, China

[3] Beijing Synchrotron Radiation Facility, Institute of High Energy Physics, Chinese Academy of Sciences, Beijing 100049, China

[4] Department of Chemistry, School of Chemistry and Biological Engineering, University of Science and Technology Beijing, Beijing 100083, China

[5] Songshan Lake Materials Laboratory, Dongguan, Guangdong 523808, China

[#] These authors contributed equally to this work.

*Corresponding author. Email: gangwang@iphy.ac.cn; hmweng@iphy.ac.cn;

Kagome metals AV$_3$Sb$_5$ (A = K, Rb, and Cs) with a V-Kagome lattice acting as a fertile platform to investigate geometric frustration, electron correlation, superconductivity, and nontrivial band topology, have attracted tremendous attention. Here we reported the structure and properties of ATi$_3$Bi$_5$ (A = Rb, Cs) family with a Ti-Kagome lattice, specifically focusing on the electronic structure and nontrivial band topology of RbTi$_3$Bi$_5$. ATi$_3$Bi$_5$ (A = Rb, Cs) is found to be non-superconducting metal with strong quasi-two-dimensional feature, moderate electron correlation, and small Pauli paramagnetism. Based on first principles calculations, RbTi$_3$Bi$_5$ is determined to be a weak topological insulator with gapless surface states along (100) plane, and the electronic band structure along (001) plane is in great agreement with experimentally observed one. In particular, the electronic properties of the RbTi$_3$Bi$_5$ monolayer can be efficiently tuned by a biaxial strain according to calculation, with its lower saddle points coming from Kagome lattice approaching the Fermi level. These results highlight ATi$_3$Bi$_5$ (A = Rb, Cs) with Ti-Kagome lattice is a new Kagome metal to explore nontrivial band topology and exotic phases.

## 1 Introduction

Kagome metals, featuring two-dimensional (2D) Kagome sub-lattice with corner-sharing triangles, have been serving as a fertile platform to investigate various electronic phenomena due to the strong geometric frustration [1] and nontrivial band topology [2]. As a model system to realize potential quantum spin liquid [3], the properties of spins on the Kagome lattice have been extensively studied [4, 5]. Owing to the unique crystal structure, Kagome metals naturally support electronic structures host Dirac points, flat bands, and saddle points, resulting in topologically nontrivial states [6-8], possible fractionalization [9, 10] and Fermi surface (FS) instabilities [11], respectively. With the combination of magnetism and topology, a 2D "Chern gap" [12] and a large anomalous Hall response [13-15] have been found in Kagome metals with the existing long-range magnetic order. By changing the filling of electrons in Kagome lattice, exotic phenomenon such as density wave orders [2, 16, 17], charge fractionalization [18, 19], and superconductivity [11, 20], could be realized. Especially when Fermi level ($E_F$) is near the saddle points, the spin or charge density wave (CDW) states would happen [11, 21-23].

Recently, the $AV_3Sb_5$ (A = K, Rb, and Cs) family with V-Kagome lattice has been discovered [24] and attracted tremendous attention due to its CDW states, superconductivity, and nontrivial band topology [25, 26]. All the $AV_3Sb_5$ compounds were found to be superconducting at low temperatures ($T_C$ ~ 2.3 K for $CsV_3Sb_5$ [24, 25], $T_C$ ~ 0.75 K for $RbV_3Sb_5$ [27], and $T_C$ ~ 0.93 K for $KV_3Sb_5$ [28]), together with an unconventional long-range CDW transition occurring at high temperatures ($T_{CDW}$ ~ 94 K for $CsV_3Sb_5$ [25], $T_{CDW}$ ~ 103 K for $RbV_3Sb_5$ [27], and $T_{CDW}$ ~ 78 K for $KV_3Sb_5$ [24]). Though without long-range magnetic order, an extraordinarily large anomalous Hall effect was observed in $KV_3Sb_5$ and $CsV_3Sb_5$ [29, 30]. By applying hydrostatic pressure or carrier doping, multiple superconducting domes can be induced in $AV_3Sb_5$, showing the competition between CDW and superconductivity [31-36]. According to density functional (DFT) calculations and angle-resolved photoemission spectroscopy (ARPES) measurements, the $AV_3Sb_5$ materials are effectively modeled as weakly correlated systems with strong quasi-2D nature [25, 37, 38]. In addition, $AV_3Sb_5$ is found to be a $Z_2$ topological metal with unconventional surface states [25, 39]. These researches highlight the materials with Kagome lattice, such as $AV_3Sb_5$, being a promising platform to explore topological superconductivity, correlated electronic states, and topological quantum computations. In the exploration of new Kagome metals, $ATi_3Bi_5$ (A = Rb, Cs) [40, 41] stands out as another family with possible superconductivity [41], orbital-selective nematic order [42] or electronic nematicity [43], and nontrivial topology [40, 41]. However, most of attention has been focused on $CsTi_3Bi_5$ rather than $RbTi_3Bi_5$, and their topological properties remain controversial and is still lack systematical investigations.

Here we report the structure and properties of $ATi_3Bi_5$ (A = Rb, Cs) family with a Ti-Kagome lattice, focusing on the electronic structure and nontrivial band topology of $RbTi_3Bi_5$. $ATi_3Bi_5$ (A = Rb, Cs) shares a similar quasi-2D structure motif with that of $AV_3Sb_5$, featuring Ti-Bi layer in Kagome lattice which is sandwiched by $A^+$ as separator. The resistivity of $ATi_3Bi_5$ (A = Rb, Cs) exhibits a metallic-like behavior without obvious anomaly down to 2 K, showing a quite large anisotropy and moderate electron correlation. A small magnetic susceptibility has been observed without any long-range magnetic order or superconducting transition in the whole measured temperature range. The electronic band structure of $RbTi_3Bi_5$ has been characterized by ARPES and fits well with DFT calculations with spin-orbit coupling (SOC), from which $RbTi_3Bi_5$ is determined to be a weak topological insulator with gapless surface states. The upper saddle points coming from Kagome lattice are proved to be far away (~ 0.8 eV) from $E_F$, whereas the lower saddle points are close (~ 0.3 eV) to $E_F$, which may result interesting states like superconductivity or CDW by changing the filling of electrons. Furthermore, due to its strong quasi-2D feature, a biaxial strain could be used to effectively modulate the electronic properties of the $RbTi_3Bi_5$ monolayer based on first principles calculations.

## 2 Experimental

**Single Crystal Growth.** $ATi_3Bi_5$ (A = Rb, Cs) single crystals were grown by a high-temperature solution method using Bi as flux. Rb/Cs chunk (99.75%, Alfa Aesar), Ti powder (99.95%, Alfa Aesar), and Bi granules (99.999%, Sinopharm) were mixed using a molar ratio of Rb/Cs : Ti : Bi =

2 : 4 : 12 in a fritted alumina crucible set (Canfield crucible set) [44] and sealed in a fused-silica ampoule at vacuum. The ampoule was heated to 1073 K over 15 h, held at the temperature for 24 h, and then slowly cooled down to 873 K at a rate of 2 K/h. At 873 K, large single crystals with size up to 5 mm ×5 mm × 0.5 mm were separated from the remaining liquid by centrifuging the ampoule. The obtained single crystals are shiny-silver plates and air-sensitive, so all manipulations and specimen preparation for structure characterization and property measurements were handled in an argon-filled glovebox.

**Structure Characterization and Composition Analysis.** X-ray diffraction data were obtained using a PANalytical X'Pert PRO diffractometer (Cu $K_\alpha$ radiation, λ = 1.54178 Å) with a graphite monochromator in a reflection mode (2$\theta$ = 5°–80°, step size = 0.017°) operated at 40 kV voltage and 40 mA current. Indexing and Rietveld refinement were performed using the DICVOL91 and FULLPROF programs [45]. Single crystal X-ray diffraction (SCXRD) data were collected using a Bruker D8 VENTURE with Mo $K_\alpha$ radiation (λ = 0.71073 Å) at 280 K for $RbTi_3Bi_5$ and a four-circle diffractometer (Rigaku XtaLAB Synergy R-DW, HyPix) with multilayer mirror graphite-monochromatized Mo $K_\alpha$ radiation (λ = 0.71073 Å) at 180 K for $CsTi_3Bi_5$. The structure was solved using a direct method and refined with the Olex2 package [46]. The morphology and element analyses were characterized using a scanning electron microscope (SEM, Phenom Prox) equipped with an electron microprobe analyzer for semiquantitative elemental analysis in energy-dispersive spectroscopy (EDS) mode. Five spots in different areas were measured on one crystal using EDS.

**Physical Property Measurements.** The resistivity, magnetic susceptibility, and heat capacity measurements were carried out using a physical property measurement system (Quantum Design, 9 T). The resistivity was measured using the standard four-probe configuration with the applied current (about 2 mA) parallel to the *ab* plane (*I* // ab) or along the *c* axis (*I* // c). Magnetic susceptibility was measured under a magnetic field (0.5 T) parallel (*H* // ab) and normal (*H* ⊥ ab) to the *ab* plane using the zero-field-cooling (ZFC) and field-cooling (FC) protocols [47].

**First Principles Calculations.** The first principles calculations were carried out with the projector augmented wave method as implemented in the Vienna ab initio simulation Package [48-50]. The generalized gradient approximation [51] of the Perdew-Burke-Ernzerhof [49] type was adopted for the exchange-correlation function. The cutoff energy of the plane-wave basis was 500 eV and the energy convergence standard was set to $10^{-6}$ eV. The 10 × 10 × 5 Monkhorst-Pack K-point mesh was employed for the Brillouin zone sampling of the 1 × 1 × 1 unit cell. The experimental crystal data were adopted to perform static calculations on $RbTi_3Bi_5$ both with and without SOC. To explore the edge states, maximally localized Wannier functions (MLWFs) for the *d* orbitals of Ti and *p* orbitals of Bi have been constructed [52, 53]. In addition, atomic SOC is added to the MLWFs based Tight Binding (TB) Hamiltonian by fitting the first principles calculations. The TB model under Wannier basis and iterative Green's function method were used to calculate the surface state [54].

**Electronic Structure Measurement.** Synchrotron ARPES and X-ray photoelectron spectroscopy measurements with various photon energies on $RbTi_3Bi_5$ were performed at beamline 4B9B of the Beijing Synchrotron Radiation Facility. In addition, a helium discharge lump (*hv* = 21.2 eV) was used as a light source during ARPES measurement. The ARPES system was equipped with a ScientaR4000 electron analyzer and the base pressure is $7 \times 10^{-11}$ Torr. The overall energy and angular resolution were 17 meV and 0.3°, respectively. RbTi$_3$Bi$_5$ single crystals were cleaved *in situ* along the (*00l*) (*l* = integer) plane.

**3 Results and Discussion**

**Crystal structure, electrical transport, and magnetization.**

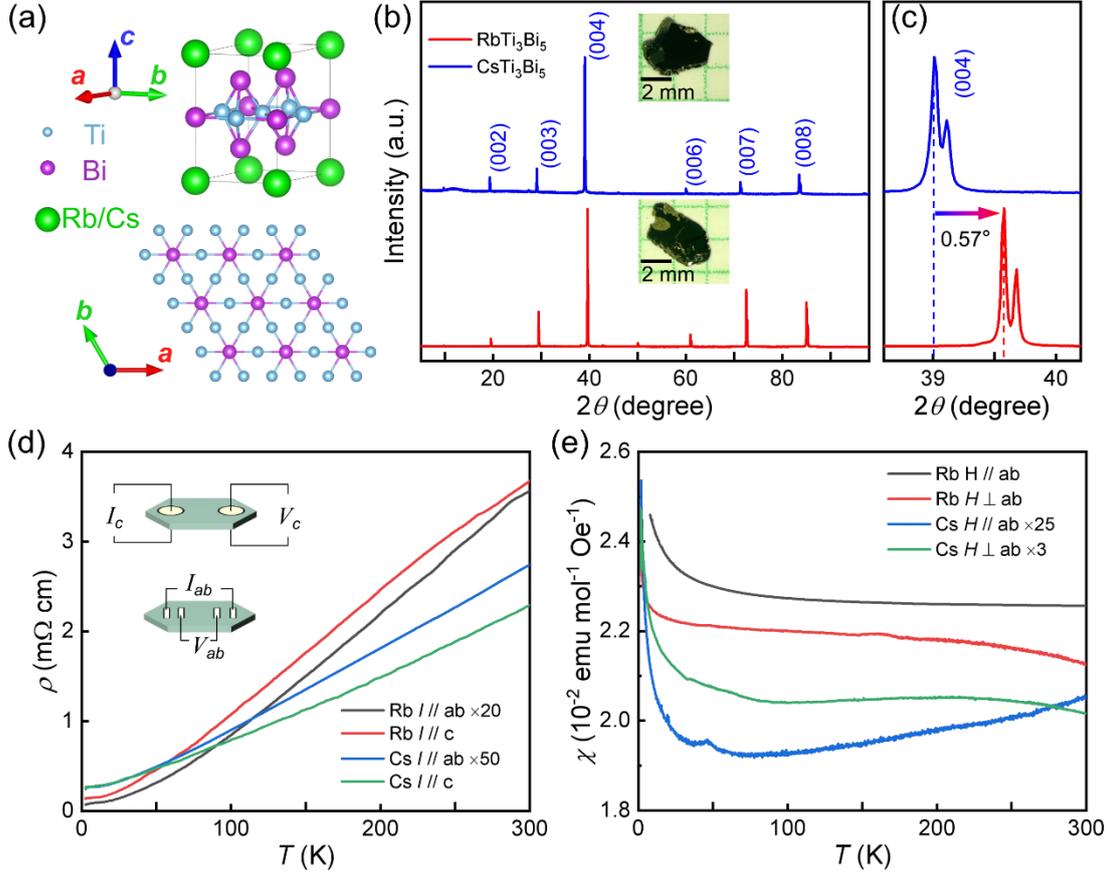

**Fig. 1. Crystal structure and physical properties of ATi$_3$Bi$_5$.** (a) Crystal structure of ATi$_3$Bi$_5$ (A = Rb, Cs) and the Ti-Bi layer with Ti-Kagome lattice. (b) X-ray diffraction patterns of as-grown ATi$_3$Bi$_5$ (A = Rb, Cs) single crystals, showing (*00l*) (*l* = integer) reflections. The insets are the optical photographs of ATi$_3$Bi$_5$ (A = Rb, Cs) single crystals. (c) Enlarged (004) diffraction peak of ATi$_3$Bi$_5$ (A = Rb, Cs). (d) Temperature-dependent resistivity of ATi$_3$Bi$_5$ (A = Rb, Cs) with *I* // ab and *I* // c. The insets show the corresponding measurement configurations. (e) Temperature-dependent susceptibility of ATi$_3$Bi$_5$ (A = Rb, Cs) with *H* // *ab* and *H* ⊥ *ab*.

The crystal structures of ATi$_3$Bi$_5$ (A = Rb, Cs) are determined based on the SCXRD data and summarized in Table SI – SIII. ATi$_3$Bi$_5$ (A = Rb, Cs) crystallizes in the hexagonal space group *P*6/*mmm* (No. 191) with $a = b = 5.8077(7)$ Å, $c = 9.1297(11)$ Å for RbTi$_3$Bi$_5$ and $a = b = 5.8079(2)$ Å, $c = 9.2400(4)$ Å for CsTi$_3$Bi$_5$, where $\alpha = \beta = 90°$ and $\gamma = 120°$. As shown in Fig. 1a, ATi$_3$Bi$_5$ (A = Rb, Cs) shares a similar quasi-2D structure motif with that of AV$_3$Sb$_5$ (A = K, Rb, and Cs), featuring Ti-Bi layer with Ti-Kagome lattice, which is sandwiched by A$^+$ acting as separator. As shown in the insets of Fig. 1b, the as-grown single crystals of $ATi_3Bi_5$ (A = Rb, Cs) are plate-like flakes with shiny metal luster, indicating a clear quasi-2D feature. The X-ray diffraction patterns of the as-grown crystals are plotted in Fig. 1b with a preferential (00$l$) ($l$ = integer) reflections. Fig. 1c shows the enlarged (004) diffraction peaks for $ATi_3Bi_5$, where the peak of $RbTi_3Bi_5$ shifts 0.57° to higher angle compared to that of $CsTi_3Bi_5$, indicating a shrinkage of distance between the structural units along the $c$ axis. The shrinkage corresponds well with the lattice parameters derived from SCXRD due to the smaller cation radius of $Rb^+$. The chemical composition is determined to be A : Ti : Bi ~ 1 :3 : 5 according to the results of EDS (Fig. S1).

With current (~2 mA) being applied in the $ab$ plane ($I$ // ab) or along the $c$ axis ($I$ // c), the resistivity of $ATi_3Bi_5$ (A = Rb, Cs) monotonically decreases with decreasing temperature (Fig. 1d), showing a metallic-like behavior without obvious anomaly down to 2 K. Both the in-plane resistivity of $ATi_3Bi_5$ (A = Rb, Cs) are much smaller than the out-of-plane resistivity, showing a quasi-2D feature. The residual-resistance ratios (*RRR*) are calculated to be 47.6 ($I$ // $ab$) and 26.2 ($I$ // $c$) for $RbTi_3Bi_5$, 8.6 ($I$ // $c$) and 10.3 ($I$ // $ab$) for $CsTi_3Bi_5$, hinting the good crystallinity of $ATi_3Bi_5$ (A = Rb, Cs) single crystals. Similiar to $AV_3Sb_5$ compounds, all the resistivity below 40 K can be well fitted using $\rho = \rho_0 + AT^\alpha$ with value of power ($\alpha$) close to 2 (Fig. S2), indicating a Fermi liquid behavior with moderate correlation between electrons [24]. Fig. 1e shows the magnetic susceptibilities of $ATi_3Bi_5$ (A = Rb, Cs) with magnetic field (0.5 T) parallel ($H$ // $ab$) and normal ($H \perp ab$) to the $ab$ plane. Above 50 K, both the samples exhibit weak Pauli paramagnetism with the small magnetic susceptibility ($< 2.4 \times 10^{-2}$ emu mol$^{-1}$ Oe$^{-1}$). The upturn at low temperature ($< 50$ K) may originate from magnetic impurities.

**Electronic structure of $RbTi_3Bi_5$.**

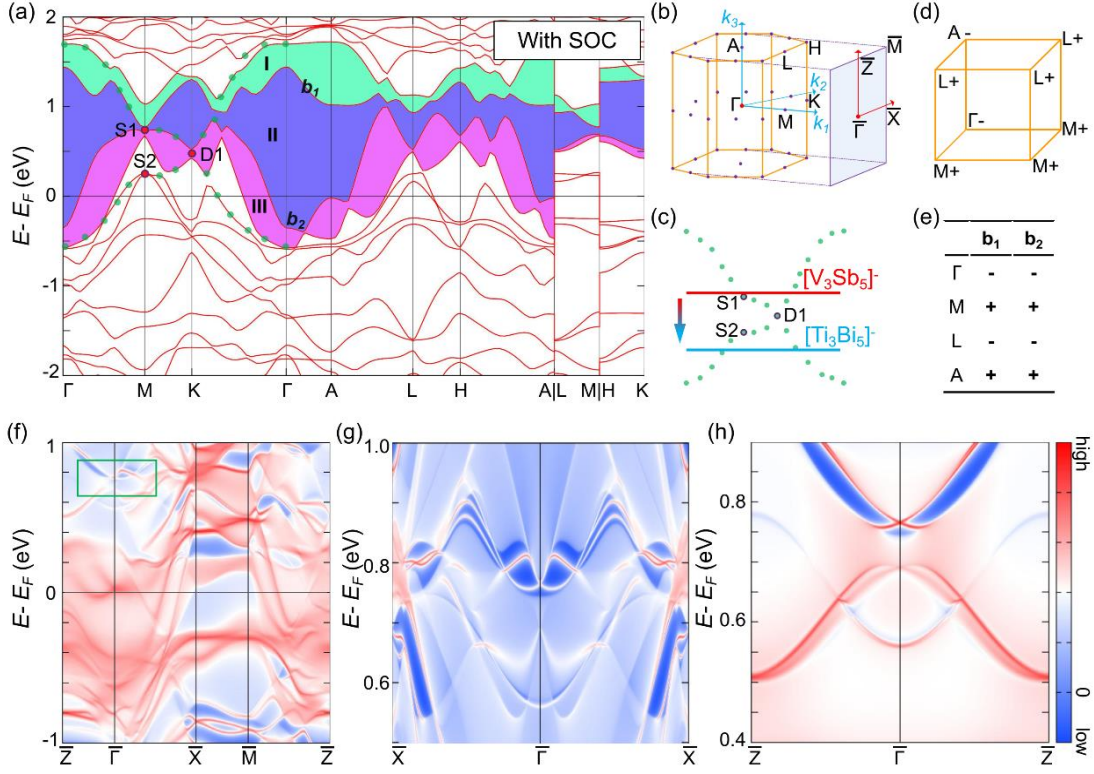

**Fig. 2. Calculated bulk and surface band structures of RbTi$_3$Bi$_5$.** (a) Calculated bulk band structure of RbTi$_3$Bi$_5$ with SOC along high symmetry lines in the first Brillouin zone (BZ). The green dots highlight the dispersions contributed mostly by the [Ti$_3$Bi$_5$]$^-$ layer with Ti-Kagome lattice, with the upper saddle points, Dirac points, and lower saddle points donated as S1, D1, and S2, respectively. The shaded areas (I, II, and III) denote the possible global gaps when considering SOC and the bands separating them are labeled as $b_1$ and $b_2$. (b) Bulk BZ and (100) surface BZ with high symmetry points and paths. (c) The schematic typical band structure and the chemical potential difference between [Vb$_3$Sb$_5$]$^-$ and [Ti$_3$Bi$_5$]$^-$ layer with Kagome sub-lattice. (d) Total parity of occupied states at eight time-reversal invariant momenta (TRIM) in bulk when considering shaded area II as the global gap. (e) Parity of the two bands $b_1$ and $b_2$ at TRIM. (f) Surface band structure of (100) plane. Zoomed-in surface band structure around $E - E_F = 0.8$ eV of (100) plane along (g) $\bar{X}$-$\bar{\Gamma}$-$\bar{X}$ and (h) $\bar{Z}$-$\bar{\Gamma}$-$\bar{Z}$.

Due to the quite similar crystal structure and physical properties, we only focus on the electronic structure and topological properties of RbTi$_3$Bi$_5$ in this work. The electronic band structure with SOC for RbTi$_3$Bi$_5$ is shown in Fig. 2a, where the bands along L-M and H-K are both fairly flat, indicating a strong quasi-2D feature. There are a number of dispersions crossing $E_F$, showing a metallic feature that is consistent with the resistivity measurement. According to the density of states (DOS) and fat band (Fig. S3), the bands near $E_F$ are mainly contributed by Ti and Bi atoms in the Kagome lattice. The typical dispersions contributed by the Ti-Kagome lattice in [Ti$_3$Bi$_5$]$^-$ layer can be clearly observed in the calculated band structures both with and without SOC, showing saddle points above $E_F$ at M (S1 around 0.8 eV and S2 around 0.3 eV) and Dirac points at K (D1 around 0.5 eV). Compared with RbV$_3$Sb$_5$ [24, 27], the number of valence electrons of RbTi$_3$Bi$_5$ decreases by three in each unit cell, resulting in the large downward shift of chemical potential (Fig. 2c). The upper saddle points (S1) and Dirac points coming from Kagome lattice in RbTi$_3$Bi$_5$ stay highly above $E_F$, making the FS instability being hard to happen, which should be the reason for the lack of CDW or superconductivity in RbTi$_3$Bi$_5$. By continuous substitution of [Vb$_3$Sb$_5$]$^-$ with [Ti$_3$Bi$_5$]$^-$ sub-lattice, $E_F$ would go through the upper saddle point S1, Dirac point D1, and lower saddle point S2 consequently with emerging interesting phases.

Nevertheless, due to the dominated influence of Kagome lattice, RbTi$_3$Bi$_5$ is anticipated to have nontrivial topological properties. Compared with the electronic band structure without SOC (Fig. S3), a number of band crossings are gapped and a global gap (shaded area I) forms far above $E_F$. In particular, some electron pockets are formed by moving down the electron band at Γ point, which gaps the band crossing along Γ-M and defines another global gap (shaded area II) near $E_F$. The Dirac point at K is also gapped and seems to form a third global gap (shaded area III). According to the location of $E_F$, the parity of high symmetry points when taking shaded area II as the global gap is calculated. As shown in Fig. 2d, RbTi$_3$Bi$_5$ is found to be a weak topological insulator having nontrivial topological invariants ($Z_2; Z_2, Z_2, Z_2$) = (0; 0, 0, 1) with the coexisting inversion symmetry and time reversal symmetry [39]. The topological properties of the other global gap (shaded area I) far above $E_F$ can be easily derived from the parity of the band separating them (Fig. 2e). By increasing the number of occupation states, like intercalation or electrical gating, RbTi$_3$Bi$_5$ can be tuned into a trivial insulator when shaded area I is considered as the global gap. According to the nontrivial topological invariants ($Z_2; Z_2, Z_2, Z_2$) = (0; 0, 0, 1), topological surface states are anticipated to exist in the planes parallel to [001] direction. In order to gain a deeper insight of the topological properties, we calculated the surface states along (100) plane of RbTi$_3$Bi$_5$. As shown in Fig. 2f, a plenty of floating surface states can be clearly observed due to the breaking of translation symmetry and reconstruction at the surface. The dispersions of surface states also show large anisotropy, with complicated dispersions along $\bar{X}$-$\bar{\Gamma}$-$\bar{X}$ but simple ones along $\bar{Z}$-$\bar{\Gamma}$-$\bar{Z}$. In the global gap (~50 meV) around high symmetry point $\bar{\Gamma}$, two brunches of surface states link the bulk valence states and conductance states (Fig. 2g, h). In particular, these two brunches of surface states form a crossing at $\bar{\Gamma}$, which should be a surface Dirac cone. These results demonstrate RbTi$_3$Bi$_5$ is a weak topological insulator with both nontrivial topological invariants and gapless surface states.

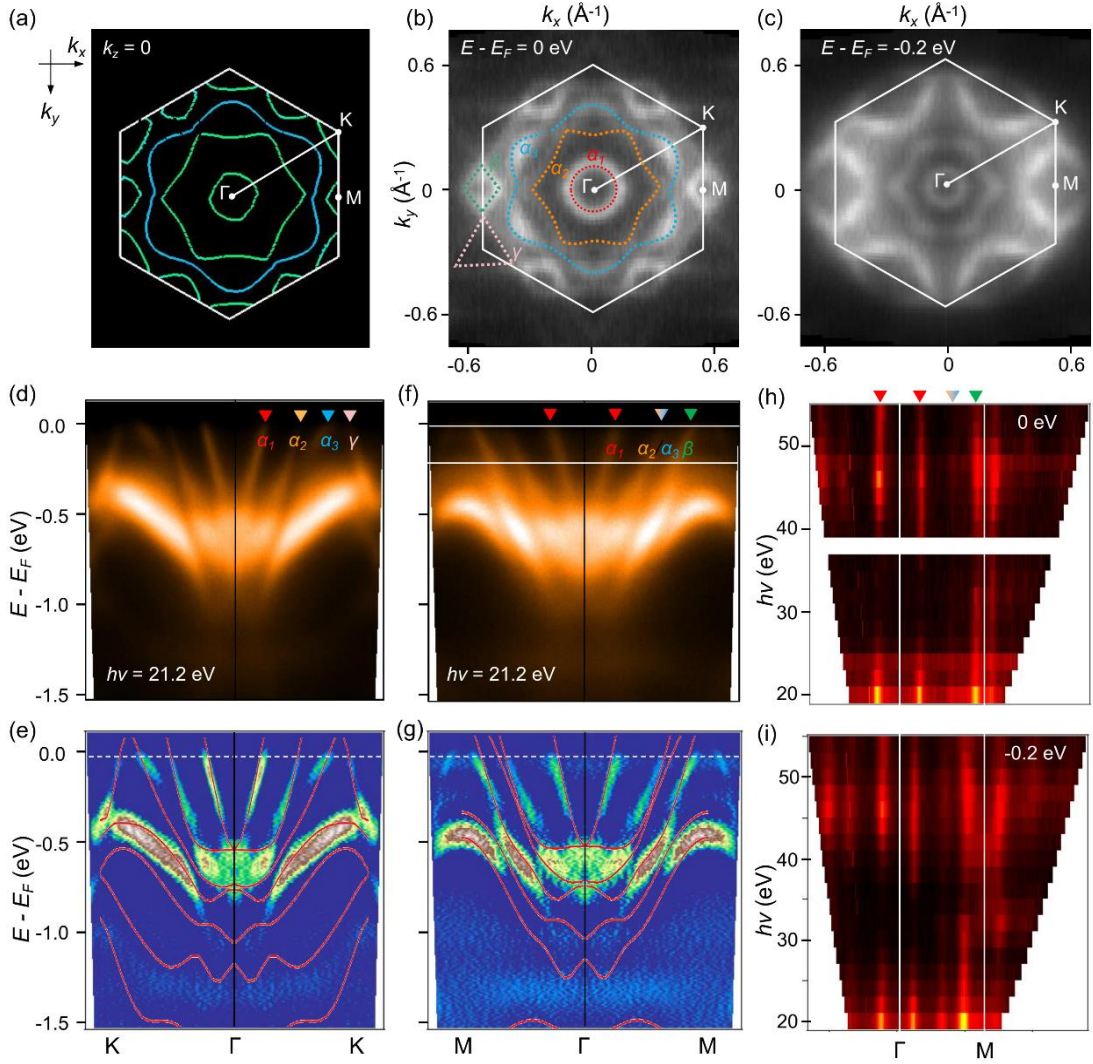

**Fig. 3. Experimentally observed band structure for RbTi$_3$Bi$_5$ and its comparison with DFT calculations.** (a) Calculated FS in Γ-M-K plane for RbTi$_3$Bi$_5$ at $E - E_F$ = 0 eV, with high-symmetry points Γ, M, and K labeled. (b) FS intensity plot in Γ-M-K plane of RbTi$_3$Bi$_5$ recorded at $hv$ = 21.2 eV, obtained by integrating the spectral weight within ±10 meV with respect to $E_F$. $\alpha_i$ ($i$ = 1, 2, 3), $\beta$, and $\gamma$ donate the multiple pockets around high-symmetry points Γ, M, and K, respectively. (c) ARPES intensity maps of Γ-M-K plane at $E - E_F$ = -0.2 eV. (d) Photoemission intensity plot along K-Γ-K with $hv$ = 21.2 eV and (e) its second derivative intensity plot. (f) Photoemission intensity plot along M-Γ-M with $hv$ = 21.2 eV and (g) its second derivative intensity plot. The arrows indicate the crossing points at $E_F$ and labeled as in (b). The red lines are the calculated band structures with SOC. Intensity plots of the ARPES data along Γ-M at (h) $E - E_F$ = 0 eV and (i) $E - E_F$ = -0.2 eV collected in a range of photon energy from 20 eV to 55 eV. The triangles indicate the corresponding crossing points at $E_F$ as in (b) and (f).

Fig. 3a shows the calculated FS in Γ-M-K plane of RbTi$_3$Bi$_5$ with SOC, which is well consistent with the measured FS of RbTi$_3$Bi$_5$ (Fig. 3b). On the *in situ* cleaved surface of (001) plane (Fig. S4), the electron pocket $\alpha_1$ around Γ point can be clearly observed, indicating the existence of large SOC interactions. Two larger hexagonal-flower-shaped electron pockets around Γ point ($\alpha_2$ and $\alpha_3$) can be further resolved, with a diamond-shaped hole pocket at M point ($\beta$) and a triangular-shaped electron pocket at K point ($\gamma$). Fig. 3c shows the ARPES intensity maps of $\Gamma$-M-K plane at $E - E_F =$ -0.2 eV, showing the shrinkage of electron pockets around $\Gamma$ or K points and the enlargement of hole pocket at M point with the increasing binding energy. Fig. 3d is the experimentally observed bands along K-$\Gamma$-K at $h\nu = 21.2$ eV, showing four crossing points at $E_F$ in a sequence of ($\alpha_1$, $\alpha_2$, $\alpha_3$, $\gamma$). As for the experimentally observed bands along M-$\Gamma$-M at $h\nu = 21.2$ eV (Fig. 3f), another four crossing points in a sequence of ($\alpha_1$, $\alpha_2$, $\alpha_3$, $\beta$) can be observed, where the intensity of $\alpha_2$ and $\alpha_3$ cannot be well resolved because they are spatially close in momentum space. By noise reduction using a machine learning process [55], a small intensity of $\alpha_2$ shows up (Fig. S5). The corresponding second derivative intensity plots (Fig. 3e and g) show remarkable agreement with the calculated electronic structure for RbTi$_3$Bi$_5$ with SOC. The experimentally observed bands can be fully recovered by calculation from $E - E_F = -1.0$ eV up to $E_F$, with the calculated $E_F$ only move down 0.05 eV. The band structure along $\Gamma$-M has been measured by varying the photon energy from 20 eV to 55 eV. As shown in Fig. 3h and i, only the change in intensity is observed at different photon energies for both $E - E_F = 0$ and -0.2 eV, indicating a strong quasi-2D feature in RbTi$_3$Bi$_5$.

Both the electron pocket $\alpha_1$ around $\Gamma$ and its well consistency with calculation indicate the existence of a strong SOC interaction, which would induce a great splitting and form a relatively large gap at M point. Thus, RbTi$_3$Bi$_5$ should be an experimentally observable weak topological insulator with surface states showing up at the global gap (shaded area II in Fig. 2a). Unlike the $Z_2$ topological insulator AV$_3$Sb$_5$, the bulk gap of RbTi$_3$Bi$_5$ is far above $E_F$ and the gapless surface states reside on the (100) plane. To experimentally observe the surface states, cleaving along the (100) plane and pumping the electrons up would be necessary. In most previous study of Kagome lattice, whether theoretically or experimentally, the upper saddle points were the focus due to the position of $E_F$ and the existence of surface states around [11, 23, 56-58]. By changing the filling of electrons and getting the upper saddle points close to $E_F$, the possible superconductivity, CDW or chiral spin order would happen. However, the lower saddle point has received little attention. Due to the large downward shift of chemical potential, the lower saddle point of RbTi$_3$Bi$_5$ is much more close to $E_F$. It would be of great interest to explore the possible new phases related to the lower saddle points by lifting $E_F$ using electron doping or strain engineering.

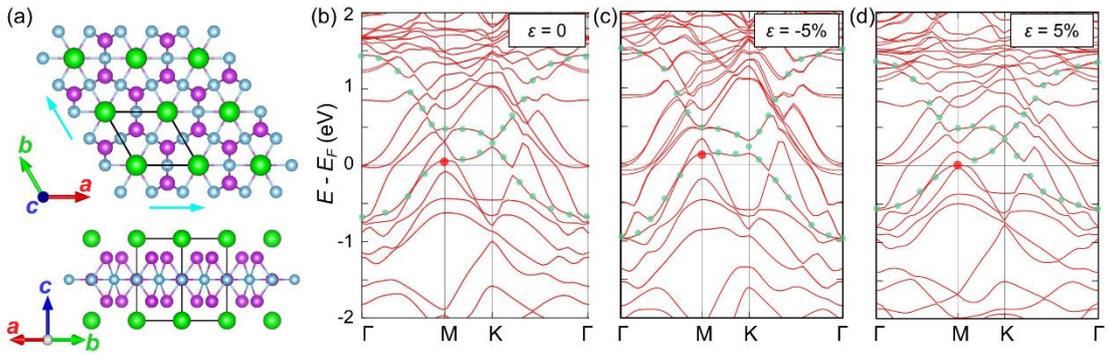

**Fig. 4. Strain engineering in RbTi$_3$Bi$_5$ monolayer.** (a) Schematic pictures of RbTi$_3$Bi$_5$ monolayer. Electronic structure of RbTi$_3$Bi$_5$ monolayer with SOC under (b) no strain ($\varepsilon = 0$), (c) -5% compressive strain ($\varepsilon = 0.95$) and (d) 5% tensile strain ($\varepsilon = 1.05$). The typical dispersions contributed by the [Ti$_3$Bi$_5$]$^-$ layer with Ti-Kagome lattice is highlighted with green dots, and the lower saddle points is marked by a red dot.

Owing to its strong quasi-2D feature, RbTi$_3$Bi$_5$ single crystal can be mechanically exfoliated into thin flakes with thickness down to 10 nm using the scotch-tape method (Fig. S6) [59], which contains several unit cells. This shows a potential for RbTi$_3$Bi$_5$ to reach the monolayer limit, making it an ideal candidate for strain engineering or electrical gating, which would be efficient in regulating its properties. Here the in-plane biaxial strain is theoretically tried to explore the effectiveness of strain engineering for modulating the electronic properties of RbTi$_3$Bi$_5$ monolayer. As shown in Fig. 4a, the RbTi$_3$Bi$_5$ monolayer is a hexagonal lattice with [Ti$_3$Bi$_5$]$^-$ layer sandwiched by Rb atoms, which is indeed stoichiometric Rb$_2$Ti$_3$Bi$_5$ and would result in an electron doping. The in-plane biaxial strain is defined as $\varepsilon = (a - a_0)/a_0 \times 100\%$, where $a$ and $a_0$ is the in-plane lattice constant of the strained and unstrained monolayer, respectively.

In RbTi$_3$Bi$_5$ monolayer, the typical dispersions contributed by the [Ti$_3$Bi$_5$]$^-$ layer with Ti-Kagome lattice can also be resolved (Fig. S7 and S8). Compared with the electronic structure of bulk RbTi$_3$Bi$_5$, the chemical potential of unstrained RbTi$_3$Bi$_5$ monolayer shifts upward for about 0.15 eV, making the lower saddle points at M (S2) closer to $E_F$ (~ 0.05 eV) (Fig. 4b). The band structure of RbTi$_3$Bi$_5$ monolayer was changed clearly with biaxial strain from −5% to 5%. With the compressive strain, the location of S2 remains above $E_F$ and goes up further (Fig. 4c), reaching 0.3 eV up to -5% compressive strain. Under the tensile strain, the location of S2 goes down and crosses $E_F$ with 5% tensile strain (Fig. 4d). By the in-plane biaxial strain, the position of the lower saddle points of RbTi$_3$Bi$_5$ monolayer can be continually tuned, which may induce interesting phases, including possible superconductivity, CDW or chiral spin order.

**4 Conclusion**

In summary, a new ATi$_3$Bi$_5$ (A = Rb, Cs) family with Ti-Kagome lattice is synthesized. ATi$_3$Bi$_5$ (A = Rb, Cs) is found to own strong quasi-2D feature, moderate electron correlation, and small Pauli paramagnetism. Based on first principles calculations considering SOC, RbTi$_3$Bi$_5$ is predicted to be a weak topological insulator with gapless surface states in its global bulk gap along the (100) plane. The experimentally observed band structure along (001) plane by ARPES fits well with the calculation and indicates the existence of strong SOC. The in-plane biaxial strain can effectively tune the position of lower saddle point coming from the Ti-Kagome lattice in its monolayer. These results show ATi$_3$Bi$_5$ (A = Rb, Cs) is a new weak topological insulator candidate and good platform to explore new phases of Kagome metals.

**Supporting information**

Tables SI-SIII show the crystal structure of ATi$_3$Bi$_5$ (A = Rb, Cs) single crystals. Figure S1 shows the SEM images and chemical composition of ATi$_3$Bi$_5$ (A = Rb, Cs) single crystals. Figure S2 shows

the power law fittings of the resistivity below 40 K. Figure S3 shows the calculated electronic structure without SOC, DOS with SOC, and fat bands both with and without SOC. Figure S4 shows the low energy electron diffraction pattern of (001) plane after *in situ* cleavage. Figure S5 shows the noise reduction of the photoemission intensity plot along M-Γ-M using a machine learning process. Figure S6 shows thin flakes of $RbTi_3Bi_5$ by mechanical exfoliation. Figure S7 shows the strain-tunable electronic structure of $RbTi_3Bi_5$ monolayer without SOC. Figure S8 shows the strain-tunable fat band of $RbTi_3Bi_5$ monolayer with SOC.

.

**Author information**

**Notes**

The authors declare no competing financial interest.

**Note added:** During the preparation of this manuscript, we recognized some unpublished preprints about $CsTi_3Bi_5$ on its superconductivity, topology, and nematicity (https://arxiv.org/abs/2209.03840, https://arxiv.org/abs/2211.12264v1, and https://arxiv.org/abs/2211.16477).

**Acknowledgements**

Y. Zhou, L. Chen, and X. C. Ji contributed equally to this work. Y. Zhou, L. Chen, and G. Wang would like to thank Prof. X. L. Chen and Prof. J. P. Hu of the Institute of Physics, Chinese Academy of Sciences for helpful discussions. This work was partially supported by the National Key Research and Development Program of China (Grant Nos. 2018YFE0202600 and 2022YFA1403900) and the National Natural Science Foundation of China (Grant No. 51832010, 11888101, Grants No. 11925408, No. 11921004, and No. 12188101), the Ministry of Science and Technology of China (Grants No. 2018YFA0305700 and No. 2022YFA1403800), the Chinese Academy of Sciences (Grant No. XDB33000000) and the Informatization Plan of the Chinese Academy of Sciences (Grant No. CAS WX2021SF-0102), and the Center for Materials Genome.

# Supporting information for
# Physical properties, electronic structure, and strain-tuned monolayer of the weak topological insulator RbTi$_3$Bi$_5$ with Kagome lattice

Ying Zhou[1, 2, #], Long Chen[1, 2, #], Xuecong Ji[1, 2, #], Chen Liu[3], Ke Liao[1, 2], Zhongnan Guo[4], Jia-Ou Wang[3], Hongming Weng[1, 2, 5, *], Gang Wang[1, 2, 5, *]

[1] Beijing National Laboratory for Condensed Matter Physics, Institute of Physics, Chinese Academy of Sciences, Beijing 100190, China

[2] University of Chinese Academy of Sciences, Beijing 100049, China

[3] Beijing Synchrotron Radiation Facility, Institute of High Energy Physics, Chinese Academy of Sciences, Beijing 100049, China

[4] Department of Chemistry, School of Chemistry and Biological Engineering, University of Science and Technology Beijing, Beijing 100083, China

[5] Songshan Lake Materials Laboratory, Dongguan, Guangdong 523808, China

[#] These authors contributed equally to this work.

[*]Corresponding author. Email: gangwang@iphy.ac.cn; hmweng@iphy.ac.cn.

# 1 Crystal structure and composition of ATi$_3$Bi$_5$ (A = Rb, Cs).

**Table SI**. Crystallographic data and structure refinement of ATi$_3$Bi$_5$ (A = Rb, Cs).

| Empirical formula | | RbTi$_3$Bi$_5$ | CsTi$_3$Bi$_5$ |
|---|---|---|---|
| f.u. weight (g/mol) | | 1278.27 | 1321.51 |
| S.G. / Z | | \multicolumn{2}{c}{P6/mmm (No.191) / 1} | |
| Unit cell parameter | a (Å) | 5.8077(7) | 5.8079(2) |
| | b (Å) | 5.8077(7) | 5.8079(2) |
| | c (Å) | 9.1297(11) | 9.2400(4) |
| | α, β (°) | 90 | 90 |
| | γ (°) | 120 | 120 |
| | V (Å$^3$) | 266.68(6) | 269.92(2) |
| $d_{cal}$ (g/cm$^3$) | | 7.928 | 8.130 |
| Refl. Collectd / unique | | 2679 / 119 | 1652 / 151 |
| $R_{int}$ | | 0.0739 | 0.0925 |
| Goodness-of-fit | | 1.537 | 1.073 |
| $R_1$ / $wR_2$ (I > 2σ(I)) | | 0.0227 / 0.0489 | 0.0251 / 0.0651 |
| $R_1$ / $wR_2$ (all) | | 0.0238 / 0.0495 | 0.0258 / 0.0655 |

**Table SII**. Atomic coordinates and equivalent isotropic displacement parameters for RbTi$_3$Bi$_5$.

| Atom | *Wyck.* | *Sym.* | x/a | y/b | z/c | Occ. | U(eq)(Å$^2$) |
|---|---|---|---|---|---|---|---|
| Rb | 1b | 6/mmm | 1.0 | 1.0 | 0 | 1.0 | 0.0338(19) |
| Ti1 | 3g | mmm | 0.5 | 0.5 | 0.5 | 1.0 | 0.0083(13) |
| Bi1 | 1a | 6/mmm | 1.0 | 1.0 | 0.5 | 1.0 | 0.0109(6) |
| Bi2 | 4h | 3m | 0.6667 | 0.3333 | 0.23386(12) | 1.0 | 0.0126(4) |

**Table SIII**. Atomic coordinates and equivalent isotropic displacement parameters for CsTi$_3$Bi$_5$.

| Atom | *Wyck.* | *Sym.* | x/a | y/b | z/c | Occ. | U(eq)(Å$^2$) |
|---|---|---|---|---|---|---|---|
| Cs | 1b | 6/mmm | 1.0 | 1.0 | 0 | 1.0 | 0.0144(6) |
| Ti1 | 3g | mmm | 0.5 | 0.5 | 0.5 | 1.0 | 0.0040(8) |
| Bi1 | 1a | 6/mmm | 1.0 | 1.0 | 0.5 | 1.0 | 0.0056(4) |
| Bi2 | 4h | 3m | 0.6667 | 0.3333 | 0.23839(7) | 1.0 | 0.0066(4) |

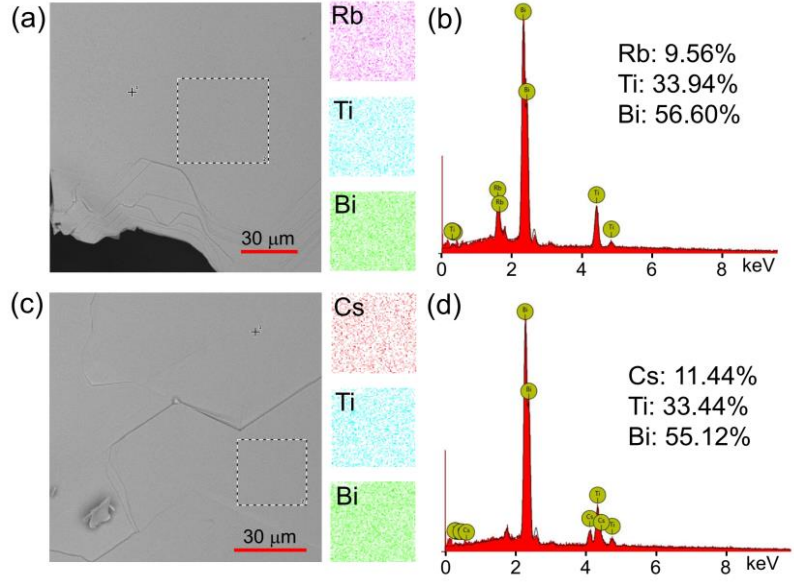

**Figure S1.** SEM images of as-grown (a) RbTi$_3$Bi$_5$ and (c) CsTi$_3$Bi$_5$ single crystals and corresponding elemental mapping. Typical EDSs and elemental composition collected on the flat clean surface of (b) RbTi$_3$Bi$_5$ and (d) CsTi$_3$Bi$_5$ single crystals.

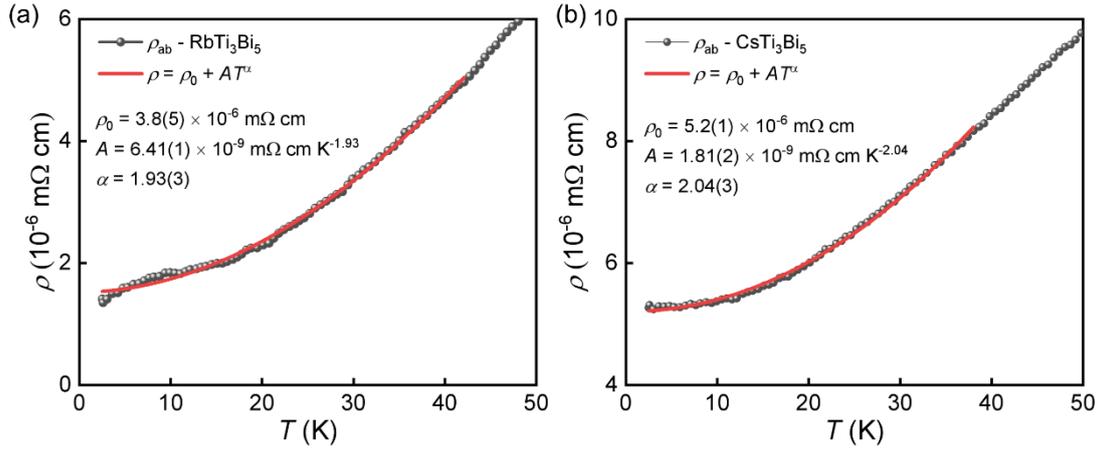

**Figure S2.** Power law fittings of the in-plane resistivity ($\rho_{ab}$) of (a) RbTi$_3$Bi$_5$ and (b) CsTi$_3$Bi$_5$ single crystals below 40 K.

**2 Electronic structure of ATi$_3$Bi$_5$ (A = Rb, Cs).**

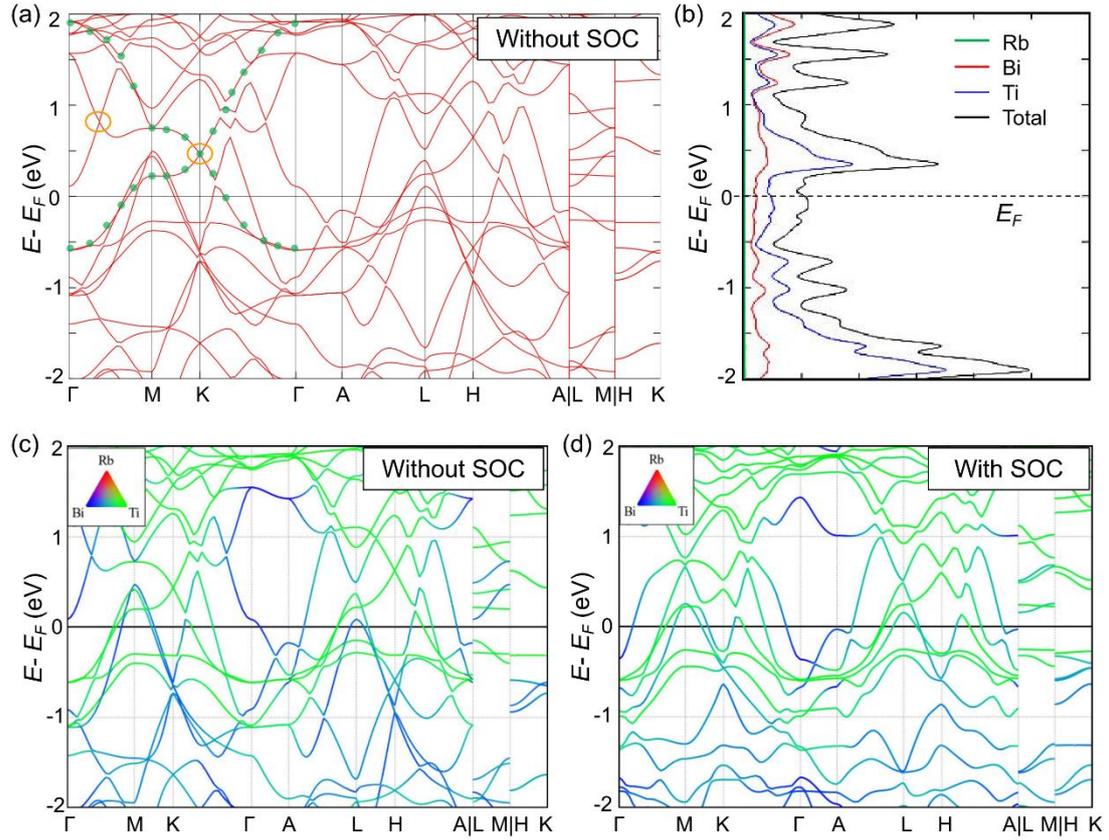

**Figure S3.** (a) Calculated band structure of RbTi$_3$Bi$_5$ without SOC along high symmetry lines in the first Brillouin zone. The green dots highlight the dispersions contributed by the [Ti$_3$Bi$_5$]$^-$ layer with Ti-Kagome lattice, and the orange circles highlight the Dirac points and band crossings. (b) Total and partial density of states near the Fermi level. Fat band of RbTi$_3$Bi$_5$ (c) without SOC and (d) with SOC.

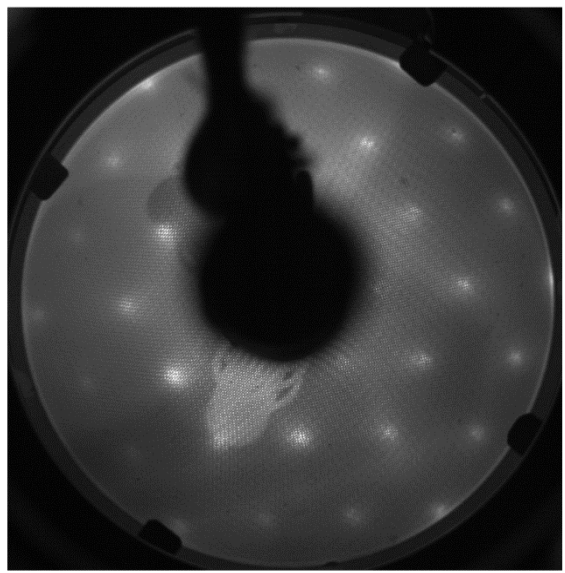

**Figure S4.** Low energy electron diffraction pattern of RbTi$_3$Bi$_5$ (001) plane after cleavage at $h\nu$ = 98 eV, showing a clear hexagonal shape.

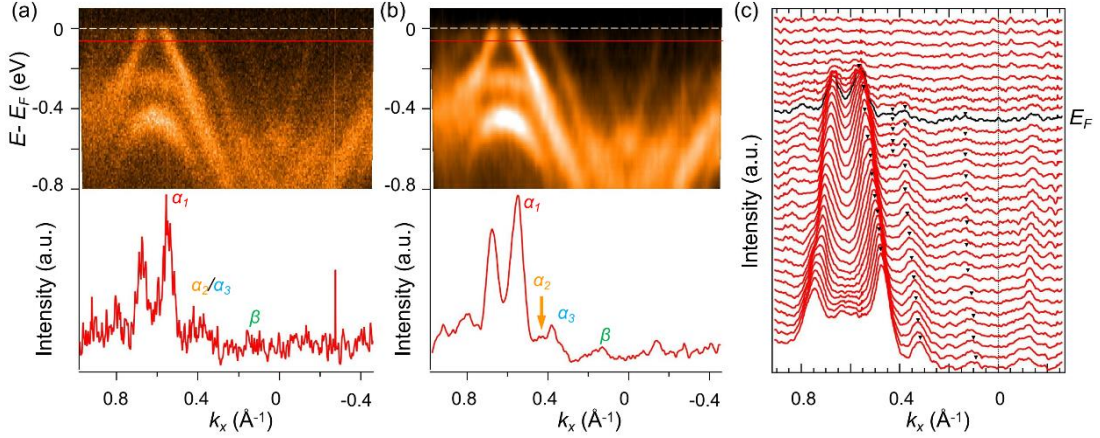

**Figure S5.** (a) Photoemission intensity plot along M-Γ-M measured with $hv$ = 30 eV and its momentum distribution curve (MDC) at $E - E_F$ = - 0.08 eV (red line). (b) Photoemission intensity plot after noise reduction of (a) using a machine learning process and its MDC at $E - E_F$ = - 0.08 eV. The white dashed line is the Fermi level and $\alpha_i$ ($i$ = 1, 2, 3), $\beta$ are the multiple pockets along M-Γ-M. (c) MDC plot of (b). The black line highlights the Fermi level and the triangles outline the shape of the dispersions.

Since the count signal collected on the angle-resolved photoemission spectroscopy detector introduces white Gaussian noise, we use the semi-supervised Noise2Noise algorithm [1], which has good ability to remove white Gaussian noise in depth learning, to suppress white Gaussian noise from raw data. Figure S5a shows the simple sum of 10 independent data acquisition of photoemission intensity plot along M-Γ-M with $hv$ = 30 eV. And the results of white Gaussian noise suppression using 10 independently collected image plots as training data are shown in Figure S5b.

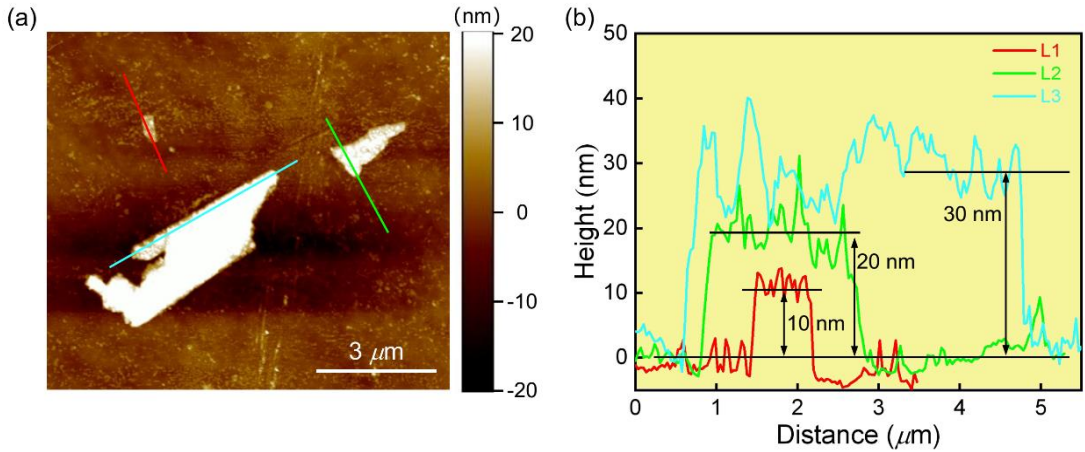

**Figure S6.** (a) Atomic force microscopy image of thin flakes of RbTi$_3$Bi$_5$ by exfoliation and (b) corresponding height profile.

Thin flakes of RbTi$_3$Bi$_5$ crystal were exfoliated using the scotch-tape method [2] and transferred onto a Si/SiO$_2$ substrate. Atomic force microscopy of thin flakes was performed using an atomic force microscope (Multimode 8.0, Bruker, USA) in a ScanAsyst mode.

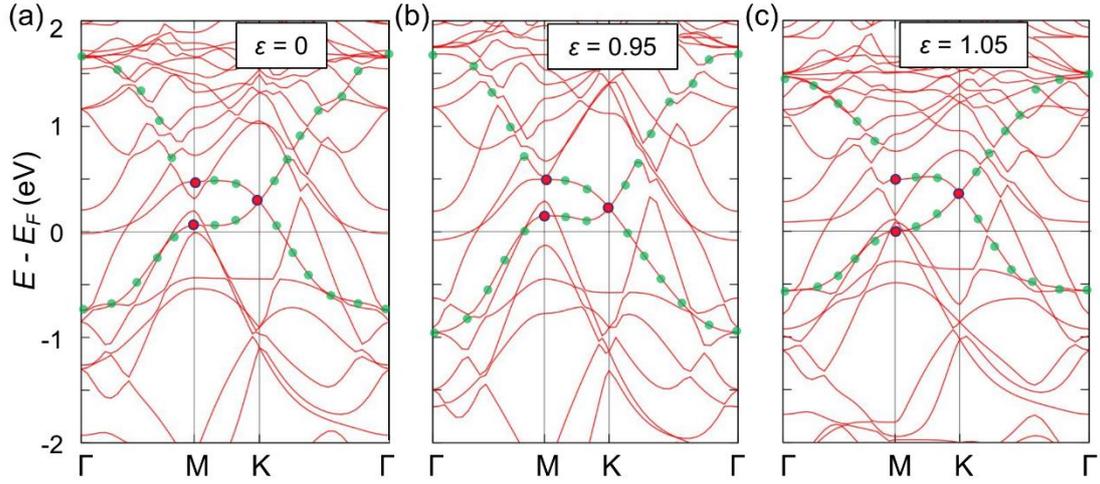

**Figure S7.** (a) Electronic structure of RbTi$_3$Bi$_5$ monolayer without SOC under (a) no strain ($\varepsilon = 0$), (b) -5% compressive strain ($\varepsilon = 0.95$) and (c) 5% tensile strain ($\varepsilon = 1.05$). The typical dispersions mostly contributed by the [Ti$_3$Bi$_5$]$^-$ layer with Ti-Kagome lattice is highlighted with green dots, and the lower saddle points is marked by a red dot.

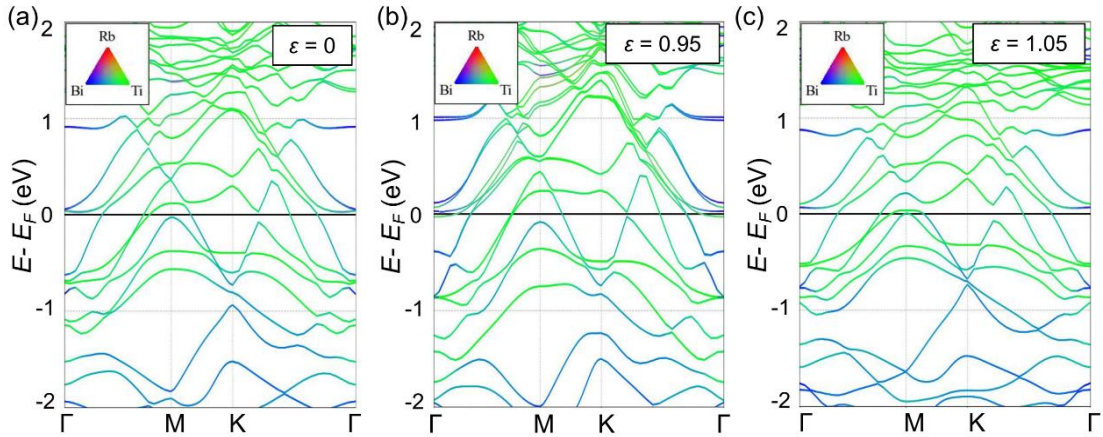

**Figure S8.** Fat band of RbTi$_3$Bi$_5$ monolayer with SOC under (a) no strain ($\varepsilon = 0$), (b) -5% compressive strain ($\varepsilon = 0.95$) and (c) 5% tensile strain ($\varepsilon = 1.05$).